\documentclass{aa}

\usepackage{savesym}
\usepackage{amssymb,amsmath}
\savesymbol{iint}
\usepackage{txfonts}
\restoresymbol{TXF}{iint}
\usepackage[T1]{fontenc}
\usepackage{graphicx}
\usepackage{epsfig}
\usepackage{multirow}
\usepackage{indentfirst}
\usepackage{natbib}

\usepackage{caption}

\newcommand{\nc}{\newcommand}
\nc{\Teff}{$T_{\rm eff}$\,}
\nc{\logg}{log\,$g$\,}
\nc{\kms}{\,${\rm km.s}^{-1}$}
\nc{\Msun}{$M_{\odot}\, $}
\nc{\Mcz}{$M_{CZ}\ $}
\nc{\vsini}{$v \sin i$}
\nc{\vrad}{$v_{\rm rad}$}
\nc{\ca}{\citealt}

\bibpunct{(}{)}{;}{a}{}{,}

\begin{document}

\title{ Shock-Induced Polarized Hydrogen Emission Lines in the Mira star omicron Ceti
\thanks{Based on spectropolarimetric observations obtained at the T\'elescope Bernard Lyot (TBL at
Observatoire du Pic du Midi, CNRS and Universit\'e de Toulouse, France). }}

\author{                        N. Fabas                                                                \inst{1}
\and                              A. L\`ebre                                                              \inst{1}
\and                              D. Gillet                                                                 \inst{2} }
\institute{                     LUPM - UMR 5299 - Universit\'{e} Montpellier II/CNRS France
\and                              Observatoire de Haute-Provence - USR 2207 - OAMP/CNRS France}

\abstract { In the spectra of variable pulsating stars, especially Mira stars, the detection of intense hydrogen emission lines has been explained by the presence of a radiative and hypersonic shock wave,  periodically propagating  throughout the stellar atmosphere. Previous observation of the Mira star omicron Ceti around one of its brightest maximum light led to the detection of a strong level of linear polarization associated to Balmer emissions, although the origin of this phenomenon is not fully explained yet.} 
{With the help of spectropolarimetry, we propose to  investigate the nature of shock waves propagating throughout the stellar atmosphere and present,  for  omicron Ceti (the prototype of Mira stars), a full observational study of hydrogen emission lines formed in the radiative region of such a shock.}
{Using the instrument  NARVAL mounted on the T\'elescope Bernard Lyot (TBL) in Pic du Midi Observatory (France), we performed a spectropolarimetric monitoring of omicron Ceti during three consecutive pulsation cycles. For this survey, the  four Stokes parameters (I for intensity, Q and U for linear polarization, and V for circular polarization) were systematically collected, with a particular emphasis on the maxima of luminosity, i.e. when a radiative shock wave is supposed to emerge from the photosphere and starts to propagate outward.}
{On hydrogen Balmer lines, over a large part of the luminosity cycle, we report clear detection of polarimetric structures in Q and U Stokes spectra (and also in V Stokes spectra but to a lesser extent). 
We report a  temporal evolution of these spectropolarimetric signatures, which appear strongly correlated to the presence of an intense shock wave responsible for the hydrogen  emission lines. 
We establish that the hydrogen lines are polarized by a physical process inherent to the mechanism responsible for the emission line formation: the shock wave itself. Two mechanisms are thus considered: a global one that implies a polarization induced by some giant convective cells located around the photosphere and a local one that implies a charge separation due to the passage of the shock wave, inducing an electrical current. Combined with the existing turbulence, this may generate a magnetic field, hence polarization.}
{}


\titlerunning{                Polarized Hydrogen Emission Lines in omicron Ceti}
\authorrunning{N. Fabas, A. L\`ebre \& D. Gillet}

\maketitle

\date{Received July 2011}

\section{Introduction}


Mira stars are cool, evolved  and  variable stars. Its prototype is omicron Ceti (as known as Mira), which is the object of this paper. On a Hertzsprung-Russell diagram, Mira stars are associated to the asymptotic giant branch (AGB). Indeed, their main-sequence mass is less than ten solar masses, which allows them to reach the AGB phase where hydrogen and helium burn in shells around a core of carbon and oxygen. They undergo radial pulsations whose origin is explained by several theories, the main one being a supposed $\kappa$-mechanism (\citealt{cox1980}), which leads to a cycle of visual light variations. Some self-excited pulsation models of Miras already exist (\ca{ireland08} and references therein), but the mechanism responsible for the pulsation has not been identified yet. The free input parameters are mass, luminosity, composition, microturbulent velocity, mixing length and turbulent viscosity. Using a non-local time-dependent theory of turbulent convection, \cite{xiong98} showed that the pulsational stability of Mira stars is governed by the dynamic and thermodynamic couplings between convection and oscillations.  \cite{xiong07} found that turbulent pressure plays a key role for  the excitation mechanism. The typical period of Mira stars' pulsation extends from 150 to 1\,000 days, the period during which the luminosity might change up to 2.5 visual magnitudes on average. It has to be noted that those stars are not as regular as, e.g., RR Lyr\ae. Therefore the duration of a cycle might vary with time.

The low surface temperature of those stars ($\sim$\,3\,000\,K) allows for the existence of molecules in the very extended atmosphere and in the circumstellar envelope, all of which are typical features of late-type stars. For the Mira stars of the M spectral type, the most present molecules are oxygen-rich ones such as H$_2$O, TiO, or CO. These molecules are at the origin of the strong molecular absorption undergone by the black-body spectrum at this temperature (\citealt{tsu1966}, \citealt{tsu1971a}, \citealt{tsu1971b}, \citealt{jorg1994}, and  \citealt{gil1988}). This absorption is such that the observed spectrum departs a lot from the corresponding black-body spectrum. Furthermore, and especially for Mira stars, strong emission has been observed on Balmer hydrogen lines, lasting during up to 80\% of the luminosity period.


In past decades, several studies - based on high resolution spectroscopic surveys - have focussed on the atmospheric dynamics of several classes of radially pulsating stars located along the instability strip: from the population II cepheid W Vir stars  (\citealt{Lebre92}) up to the RV Tauri stars (\citealt{Lebre91}), the Mira stars (\citealt{Alvarez00} and \citealt{Alvarez01}), and the post-AGB stars (\citealt{Lebre96}). It is now admitted  that radially pulsating variable stars host radiative hypersonic shock waves that propagate periodically (the period is hardly changing with time) throughout  the stellar atmosphere.

Two main  properties of these  shock waves  have been investigated: (i) their radiative nature, generating emission lines formation process (hydrogen and helium lines, see \citealt{gilletal1983} ;  fluorescent lines, see \citealt{will1976}), and (ii) their hypersonic nature, from an estimation of their propagation velocity that always appears higher than the sound velocity in such media (several Mach numbers ; see \citealt{Gillet98} and references therein). Their structure has also been investigated (see \citealt{fadgil2004} and references therein), and three main areas constitute those radiative shock waves. First of all, the radiative precursor is the area before the shock where the matter has not yet been affected by the raising wave compression. Then comes the shock front. In a length of only several mean free paths, a part of the kinetic energy is brutally converted into thermal energy, which creates strong temperature, pressure, and velocity gradients. At the end, the radiative wake is located behind the front. There, the partially ionized matter gets cooler and recombines. The intense emission lines observed in the spectra of radially pulsating variable stars, such as Mira stars, are supposed to be formed just behind the shock front (\citealt{gilletal1983}) in the so-called \textit{radiative} wake of the shock.

 These studies have also related the shock wave propagation phenomenon in the atmospheres of variable stars to their pulsating process (which is the $\kappa$-mechanism for the large majority of these objects) or to more chaotic behaviour (for RV Tauri and Post-AGB stars, for example, \citealt{Barthes2000}).  These works have mainly established that shock waves, which are connected to the pulsation mechanism,  are emerging from the photosphere once a cycle, around the maximum light,  and are thus propagating periodically throughout stellar atmosphere, ruling its dynamics, and extending it to several tens of solar radii,  progressively generating, and shaping a circumstellar envelope around coolest objects.

 However, to date, the magnetic nature of these shock waves has never been identified or deeply investigated, while since the 60s, theoretical studies of shock waves propagating throughout gaseous media of atmospheric density predict there is an electric field (therefore a magnetic field) in the post-shock region  (see for instance: \citealt{Jaffrin65} ; \citealt{Lu74} ; \citealt{Vidal95}).  Moreover, because of the complexity of the problem, this magnetic nature has never been introduced in shock structure or shock propagation modelling. Also, its impact on astrophysical processes, such as radial pulsation or chaotic behaviour, atmospheric dynamics, circumstellar envelope formation and structure, mass loss, and grain formation, has never been investigated so far.

 In the late 70s, narrow- and broad-band polarimetric observations were performed on cool evolved stars hosting circumstellar envelopes in order to investigate the grain distribution in stellar environments. With this technique, in a  study devoted to the  one-year-period variable Mira star, omicron Ceti,  \cite{mlc1978} were the first to 
detect a strongly linear polarized flux in the emission lines of the Balmer series. Based on just a single observation limited to the maximum luminosity, the polarization in the hydrogen lines (from $H\beta$ to H10) was noticed as two to three times greater than the polarization measured in the nearby continumm. To date, this observational fact is still an enigma. It has also never been reproduced on any Mira star or on another pulsating star, or even characterized during a complete pulsating cycle. Moreover, recent modelling of the linear polarization of optical emission in red giants (\citealt{Fadeyev07}) which investigates light scattering by dust grains in axisymmetric circumstellar envelopes, has even failed to explain such a high polarization rate.

Recently, a new generation of instruments has been made available:  spectropolarimeters.  Over a large spectral domain, they combine high-resolution spectroscopy and  information in polarized light, and they open new observational windows for astrophysical objects, mainly focussing on stellar magnetism. With these instruments, it is now possible to collect spectral information through the four Stokes parameters (V, characterizing circular polarization ; Q and U, characterizing linear polarization ; I, the classical spectrum).

\cite{Harrington09} present spectropolarimetric observations of Post-AGB stars and RV Tauri stars and report, for most of these objects,  the detection of linear (Stokes Q and Stokes U) polarimetric signatures associated to the $H\alpha$ lines. In most objects, the morphology of the polarization feature presents short timescale variability that suggests the presence of absorbing polarizing gas  close to the star, according to the optical pumping model described by \cite{Kuhn07}. Stong mass loss, progressively shaping circumstellar environments, is very common in these stars, and the shock waves propagating throughout their atmosphere and inducing ballistic motions has also been reported (\citealt{Lebre91}, \citealt{Lebre96}).

It is thus now possible for the very first time to estimate the linear and the circular polarization evolution in the emission lines of Mira stars. As we are convinced that the hydrogen emission lines are polarized by a physical process inherent to the mechanism responsible for the emission line formation (i.e, the shock wave itself), a spectropolarimetric survey performed during the cycle of pulsation of Mira stars appears to be a good tool for tackling this problem. In section 2, we present our observations of omicron Ceti as collected with the NARVAL spectropolarimeter. In section 3, we interpret the obtained results in term of shock wave characterization. In section 4, we draw a conclusion from this spectropolarimetric analysis.

\section{Spectropolarimetric monitoring of omicron Ceti}

\subsection{Observations}

We performed a spectropolarimetric monitoring along three consecutive pulsation cycles of the bright star omicron Ceti, the prototype of oxygen-rich Mira stars. During one of its typical cycles of luminosity (with a mean period of pulsation of 331 days), the visual magnitude of omicron Ceti presents an average variation amplitude of eight and its spectral type varies from M5e to M9e (see General Catalogue of Variable Stars and AAVSO website). Omicron Ceti has the peculiarity of hosting a companion (VZ Ceti) at $\sim$70 A.U. believed to be either a white dwarf (\citealt{Karovska97}, \citealt{Karovska06}) or a main sequence star (\citealt{Ireland07}). We consider it  to be too far to have any influence on the dynamics of the lower atmosphere of omicron Ceti.

The spectropolarimetric data come from the NARVAL instrument mounted on the T\'elescope Bernard Lyot (TBL) in the Pic du Midi Observatory, France. The polarimeter is made with optical elements (a series of half-wave, quarter-wave, and half-wave rhombs) that analyse the circular or linear polarization state of a signal (\citealt{donatial1997}). The processed signal is thus transmitted in an achromatic way via an optical fiber to the cross-dispersion spectrometer doing the spectral analysis over a very wide part of the spectrum (from 375 to 1\,050\,nm) and over 40 orders. The polarimetric mode gives us a spectral resolution of 68\,000, and the velocity bin is then $\Delta$v\,$\sim\,$4.4\,km.s$^{-1}$.

\begin{figure}[h]
\caption{Light curve of omicron Ceti (from AAVSO website) for the three cycles of the survey. Each blue cross is one spectropolarimetric observation of our survey.}
\label{cdl}
\includegraphics[scale=0.5]{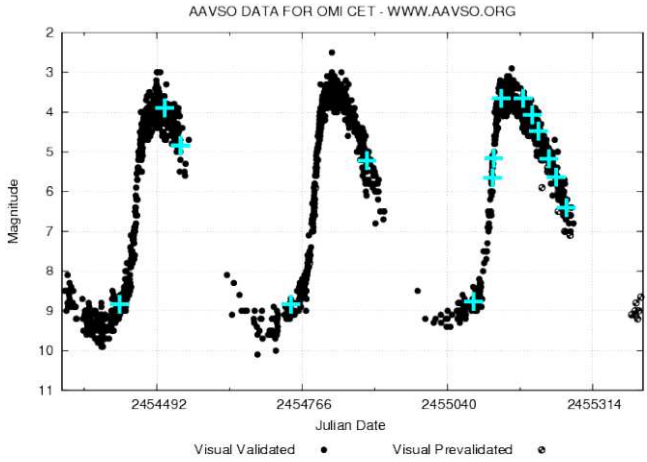}
\end{figure}

The monitoring was from the 5 September 2007 to the 10 February 2010, covering three luminosity cycles of omicron Ceti. To provide phases during these cycles, we used as heliocentric julian date of reference at luminosity maximum (see AAVSO website): $\mathrm{HJD}_\mathrm{r}$ = $2\,454\,150$. The phase $\varphi$ is then the ratio between $\mathrm{HJD}_\mathrm{o}-\mathrm{HJD}_\mathrm{r}$ and the period (see Fig. \ref{cdl} and Table \ref{tab1}).

\begin{table}[t]

\caption{Journal of our spectropolarimetric observations of omicron Ceti with Narval.}
\label{tab1}
\begin{minipage}{\linewidth}
\begin{tabular}{|c|c|c|c|c|c|}
\hline
Date&Phase&HJD&Exp.&Stokes&S/N\footnote{The signal-to-noise ratio refers to the Stokes I parameter at  412 nm and 647 nm, respectively.}\\
(dd/mm/yyyy)&$\varphi$&(2454000+)&time(s)&param.&\\
\hline

\multirow{3}{*}{05/09/2007}&\multirow{3}{*}{0.58} & 348.63529&1200&V&29/277\\
&&348.65190&1200&Q&27/268\\
&&348.66850&1200&U&31/278\\
\hline

\multirow{9}{*}{20/01/2008}&\multirow{9}{*}{1.00} &486.23594&120&Q&139/704\\
 &&486.24927&120&U&134/706\\
 &&486.25322&120&V&136/707\\
 &&486.25732&120&Q&140/695\\
 &&486.26126&120&U&135/690\\
 &&486.26520&120&V&125/640\\
 &&486.27026&120&Q&142/723\\
 &&486.27420&120&U&144/741\\
 &&486.27817&120&V&134/681\\

\hline

\multirow{3}{*}{10/02/2008}&\multirow{3}{*}{1.06}&507.25895&120&Q&181/929\\
&&507.26289&120&U&175/931\\
&&507.26683&120&V&179/945\\

\hline

\multirow{3}{*}{29/08/2008}&\multirow{3}{*}{1.68}&708.64006&1200&Q&22/302\\
&&708.65652&1200&U&21/288\\
&&708.67299&1200&V&23/304\\

\hline

\multirow{6}{*}{26/02/2009}&\multirow{6}{*}{2.23}&889.26324&120&Q&132/525\\
&&889.26716&120&U&122/506\\
&&889.27113&120&V&130/538\\
&&889.27522&120&Q&133/514\\
&&889.27914&120&U&132/536\\
&&889.28309&120&V&100/415\\

\hline

24/07/2009&2.67&1037.63799&520&V&16/223\\

\hline

\multirow{3}{*}{23/09/2009}&\multirow{3}{*}{2.86}&1098.49282&400&Q&15/212\\
&&1098.50000&400&U&17/208\\
&&1098.50718&400&V&15/204\\

\hline
\multirow{3}{*}{03/10/2009}&\multirow{3}{*}{2.89} &1108.44130&400&Q&28/331\\
 &&1108.44847&400&U&32/367\\
 &&1108.45565&400&V&28/324\\

\hline

\multirow{3}{*}{25/10/2009}&\multirow{3}{*}{2.95} &1130.50330&120&Q&109/575\\
 &&1130.50725&120&U&102/537\\
 &&1130.51119&120&V&104/549\\

\hline

\multirow{3}{*}{24/11/2009}&\multirow{3}{*}{3.05} &1160.32884&40&Q&57/327\\
 &&1160.33185&40&U&55/331\\
 &&1160.33487&40&V&56/337\\

\hline

\multirow{3}{*}{10/12/2009}&\multirow{3}{*}{3.09} &1176.35850&160&Q&135/695\\
 &&1176.36289&160&U&138/700\\
 &&1176.36728&160&V&150/759\\

\hline

\multirow{3}{*}{20/12/2009}&\multirow{3}{*}{3.12} &1186.26722&28&Q&61/350\\
 &&1186.27008&28&U&59/334\\
 &&1186.27294&28&V&58/329\\

\hline

\multirow{3}{*}{06/01/2010}&\multirow{3}{*}{3.18} &1203.35451&40&Q&41/310\\
 &&1203.35750&40&U&42/311\\
 &&1203.36053&40&V&40/285\\

\hline

\multirow{3}{*}{18/01/2010}&\multirow{3}{*}{3.21} &1215.30800&160&Q&116/735\\
 &&1215.31242&160&U&112/715\\
 &&1215.31681&160&V&111/707\\

\hline

\multirow{3}{*}{10/02/2010}&\multirow{3}{*}{3.28} &1238.27729&520&Q&66/495\\
&&1238.28585&520&U&47/355\\
&&1238.29443&520&V&44/344\\

\hline

\end{tabular}
\end{minipage}
\end{table}

\subsection{Data treatment}

The Libre-ESpRIT software (\citealt{donatial1997}) processing data collected by NARVAL allows us to obtain polarimetric spectra. The outputs of Libre-ESpRIT process are the four Stokes parameters divided by a pseudo-continuum:
\begin{itemize}
\item{the intensity \Large{$\frac{I}{I_{c}}$},}\\
\item{the linear polarization 

\begin{equation*}
\mathrm{q}=\frac{Q}{I_{c}}=\frac{I_{0^{o}}-I_{90^{o}}}{I_{c}}
\end{equation*}
\begin{equation*}
\mathrm{u}=\frac{U}{I_{c}}=\frac{I_{45^{o}}-I_{135^{o}}}{I_{c}},
\end{equation*}
}\\
\item{the circular polarization
\begin{equation*}
\mathrm{v}=\frac{V}{I_{c}} = \frac{ I_{\circlearrowleft}-I_{\circlearrowright} } {I_{c}}
\end{equation*}

(I$_{\circlearrowleft}$:left-handed circular polarization, I$_{\circlearrowright}$: right-handed circular polarization).

}
\end{itemize}

From this information, we can deduce the rate and the angle of linear polarization (see \citealt{Bagnulo09}):

\begin{equation*}
\frac{p_{lin}}{I_{c}} = \frac{\sqrt{Q^{2}+U^{2}}}{I_{c}},\;\theta=\frac{1}{2}\mathrm{arctan}\left(\frac{U}{Q}\right)+C\\
\end{equation*}
\begin{equation*}
\mathrm{with}\; \left\{
\begin{array}{lcl}
C=0^{\circ}&\mathrm{if}&Q\;\mathrm{and}\;U>0\\
C=90^{\circ}&\mathrm{if}&Q<0\\
C=180^{\circ}&\mathrm{if}&Q>0\;\mathrm{and}\;U<0\\
\end{array}
\right.
\end{equation*}

The Libre-ESpRIT software automatically corrects a pseudo-continuum. It deduces a black-body continuum from the observed spectrum by which this spectrum is divided. This feature is quite a problem in our case because the spectra of cool Mira stars are very affected by molecular absorption. Moreover, because of the pulsational mechanism, the effective temperature presents a strong variation in the luminosity cycle (as well as the associated black-body radiation). However, the emission features that we are analysing are often very intense (hence much higher than the local continuum). In the blue part of the spectrum, where most of the Balmer lines are, the molecular absorption is less, and the continuum change is assumed to be negligible with respect to the extent of the Balmer lines. However, we chose to use non-normalized data and a pseudo-continuum defined by ourselves. A polynomial fit of second order is calculated on the pseudo-continuum around the line and between -200\kms and 200\kms. Then, we normalize the data with this fit.

The polarization rates given in our study might reach unexpectedly high values (several dozen percent). One could say that no polarizing mechanism may be able to polarize light as efficiently, but it has to be borne in mind that this rate is given as a percentage of the pseudo-continuum intensity, i.e. not as a percentage of the total intensity (where the total intensity meant as the sum of the pseudo-continuum intensity and the intensity in the strong emission line), which is what is usually given in theoretical works. This is the first time (except in \citealt{mlc1978}) that a polarimetric study is presented by analysing such intense emission.

Compared to the work presented in \citealt{mlc1978} and also devoted to omicron Ceti, our study is based on much more polarimetric observations collected at various phases. Since we present the evolution of the intensity and polarization with respect to the pseudo-continuum, those observations made through time are thus believed to be comparable.
 
A threshold of detectability can be set at $10^{-4}$ (J. Morin, private communication). Also, a limiting constraint is the crosstalk between different Stokes parameters owing to instrumental problems. It has been measured to be no more than 1\% (see NARVAL website). Knowing that, we chose 3\% as a reasonable limit for polarimetric signatures to be considered as detections.

\subsection{Spectropolarimetric data around Balmer lines}

Figures \ref{fig:spec1} to \ref{fig:spec4} present the polarimetric data (in $H\gamma$ and $H\delta$) of omicron Ceti for the best-sampled cycle of our survey (the third cycle). 
Our analysis is focussed on those two Balmer lines, which are the least affected by TiO absorptions. In Figs \ref{fig:spec1} and \ref{fig:spec3}, the spectra of the four Stokes parameters I, Q, U, and V are displayed. The abscissa is taken as velocity in the stellar rest frame (omicron Ceti's radial velocity: V$_{*}$=63.8\kms, \citealt{Wilson63}),  and the spectra were convolved with a Gaussian curve of FWHM\,=\,4\,\kms to slightly reduce noise and improve the quality of the data. In Figs \ref{fig:spec2} and \ref{fig:spec4}, we also show u(q), which can  be read as a polar graph with the linear polarization rate $p_{lin}$ as the radial value and $2\theta$ as the angle.

\subsubsection{$H\gamma$}

\begin{figure*}[h!]
\begin{center}
\includegraphics[scale=0.6]{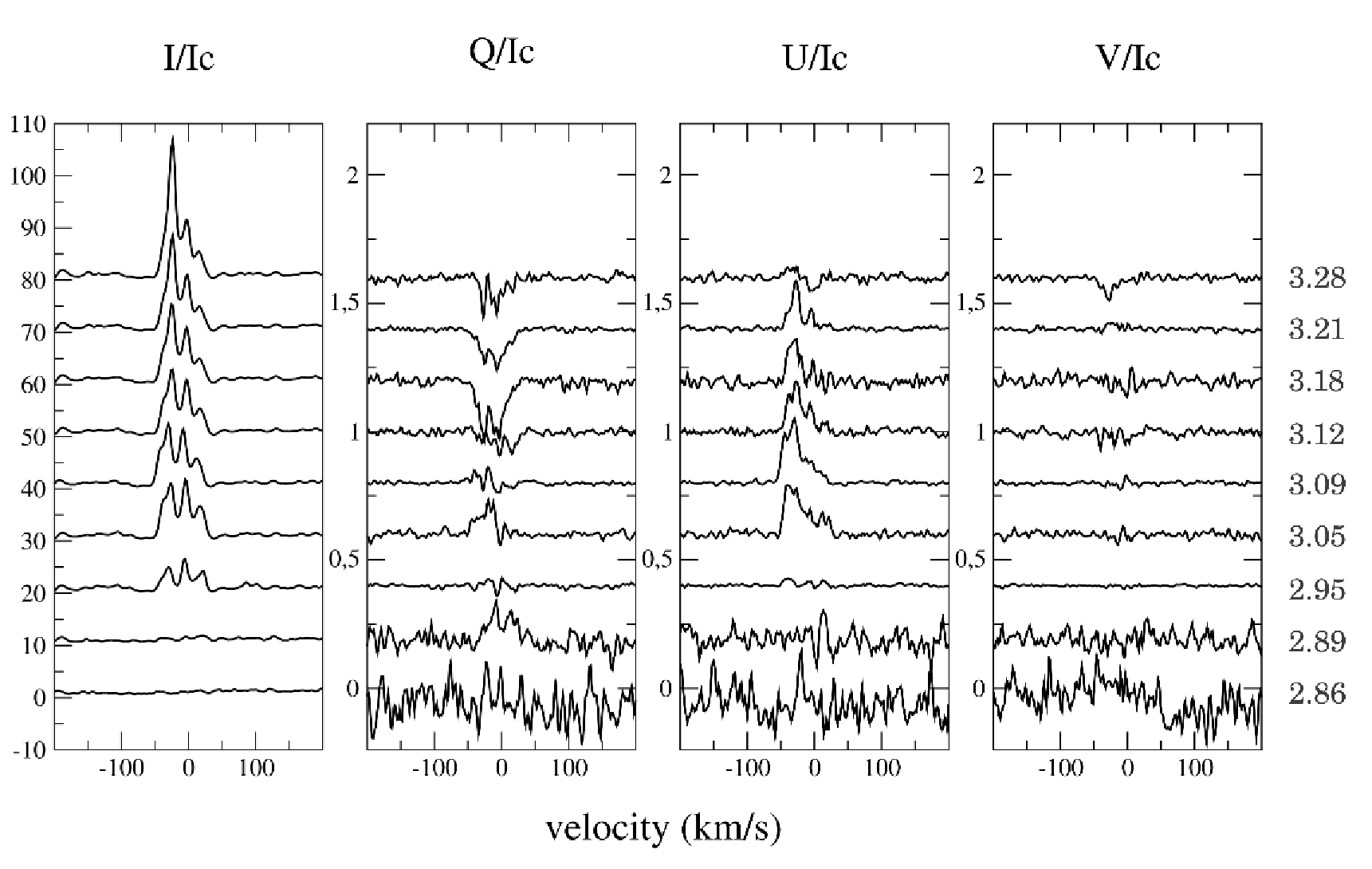}
\caption{Hydrogen Balmer lines signatures of the third cycle of our monitoring of omicron Ceti in $H\gamma$ (full Stokes). Successive signatures are artificially offset vertically for legibility. Phases are indicated on the right (see Table \ref{tab1}).}\label{fig:spec1}
\end{center}  
\end{figure*}

\begin{figure*}[h!]
\begin{center}
\includegraphics[scale=0.4]{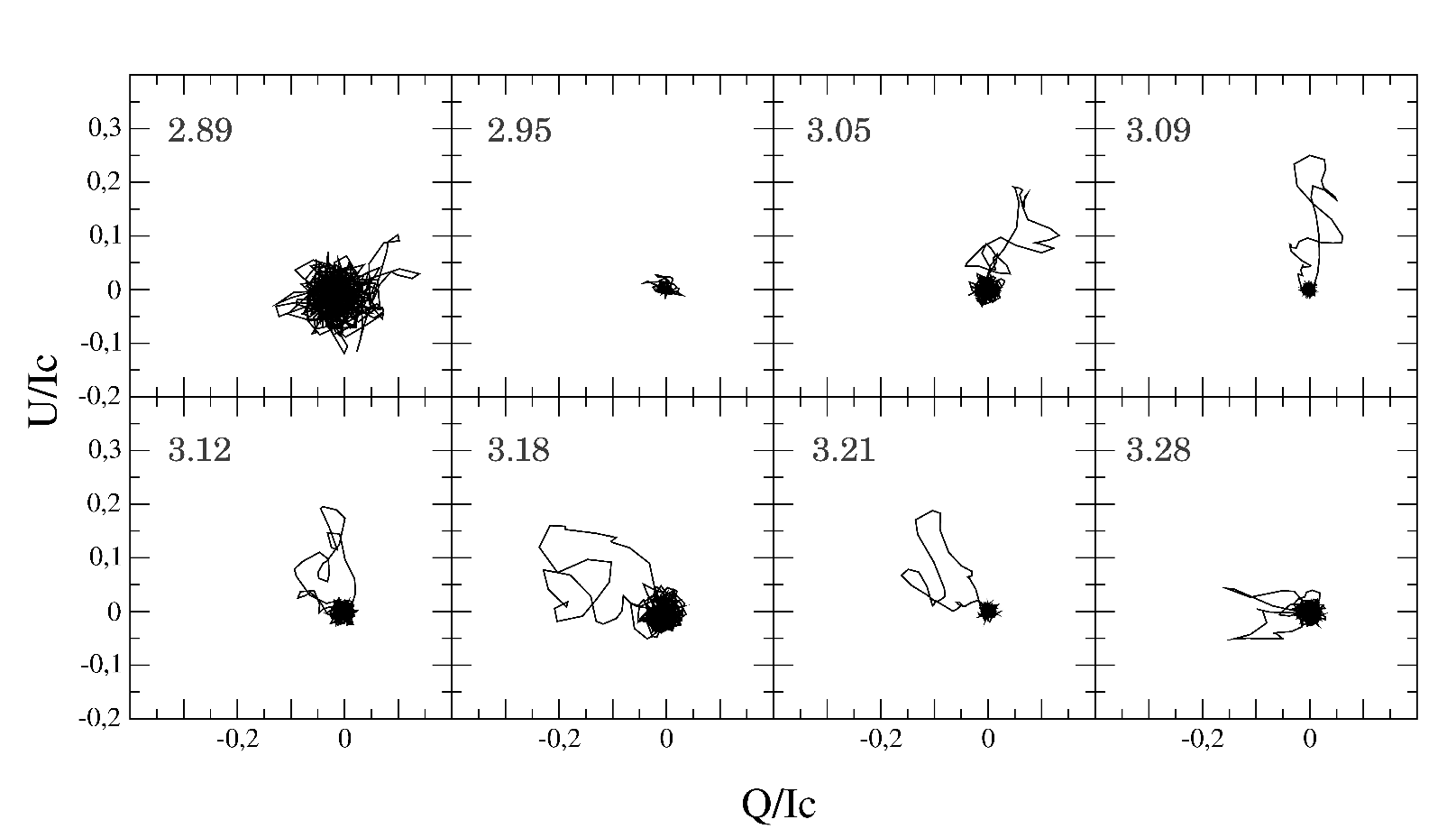}
\caption{Hydrogen Balmer lines signatures of the third cycle of our monitoring of omicron Ceti in $H\gamma$ (u(q)). Phases are indicated in the topleft of each frame (see Table \ref{tab1}).}\label{fig:spec2}
\end{center}  
\end{figure*}

From Figs \ref{fig:spec1} and \ref{fig:spec2}, devoted to the data around the Balmer line  $H\gamma$,  the emission appears at $\varphi$=2.95, hence slightly before the maximum light.  During the full duration of our monitoring, the $H\gamma$ emission line increases steadily and displays some well known features (\citealt{Joy47}, \citealt{gil1988}, \citealt{Castelaz00}).  Its time evolution can be related to the presence of a  shock wave  emerging from the photosphere just before the maximum light and starting to propagate outward. While it first encounters a medium of decreasing density in the lower part of the atmosphere, the shock thus accelerates, before being progressively dimmed in the highest atmospheric layers because of the energy lost through radiation. While it is doing so, the column densities of absorbants (located above the shock's front) decreases slightly. Those molecular absorptions are shaping the emission line to leave three peaks that can be easily identified from $\varphi$=2.95 to $\varphi$=3.28. The most redshifted peak is constant (in intensity) through time, in contrast to the two others. The central peak evolves  quite rapidly around the maximum light and then remains constant in intensity until  the end of our observations. The  bluest peak develops progressively and reaches $\varphi$=3.28 at an intensity that is five times higher than the one displayed at $\varphi$=2.95. On its blue edge,a shoulder is also present and  disappears around $\varphi$=3.21. Each of these peaks is globally redshifted with time with hardly any change in the width. Throughout our observations, the total width of this emission is conserved through time at 90 \kms.

When the intensity starts to develop, the Stokes parameters q and u both present strong signatures. Stokes q has a positive value at phase $\varphi$=2.89, 3.05. It becomes zero and then negative through time, reaching its full amplitude at $\varphi$=3.18. All the signatures in q spread over a velocity width equivalent to the one of the intensity emission and the features often present a line reversal at their centre. The Stokes u signature remains positive throughout our monitoring. It reaches its maximum amplitude at  $\varphi$=3.09 and remains quite constant in shape and in amplitude until $\varphi$=3.12. After the maximum light, the signatures on u present a strong change in shape, because strongly asymmetric and because clearly becoming narrower with time. In a less definite way, Stokes v signatures are very weak and present structure variations through time. While the radiation is clearly getting linearly polarized during the shock's propagation (from $\varphi$=3.05 to $\varphi$=3.28), circular polarization is hardly detected except at the very end of our monitoring.

The variation in the angle of linear polarization is quite obvious when looking at the u(q) plots (Figure \ref{fig:spec2}). At $\varphi$=3.05, the first phase of definite polarization,  $\theta$ is around $20^{\circ}$. Afterwards the direction of linear polarization rotates counterclockwise to reach $90^{\circ}$ at $\varphi$=3.28. After the maximum light, at phase $\varphi$=3.05, the linear polarization rate reaches 0.2  and stays around this value except for the last observation at $\varphi$=3.28, where it decreases to 0.1. However, noticeable is that, when the intensity emission is still growing, the linear polarization rate decreases after its maximum. Whatever is polarizing light works less efficiently at this specific moment.  


\subsubsection{$H\delta$}

\begin{figure*}[h!]
\begin{center}
\includegraphics[scale=0.6]{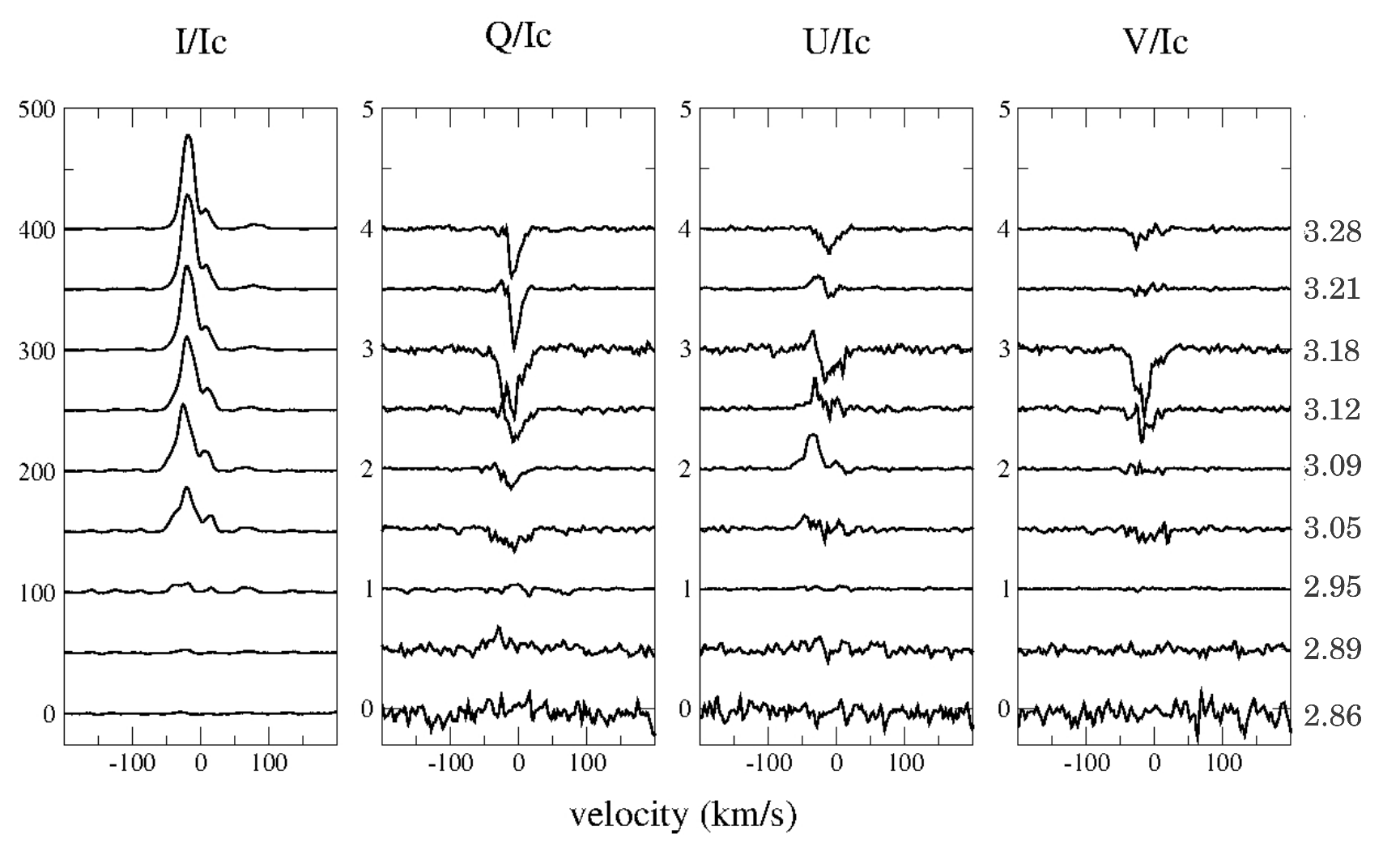}
\caption{Hydrogen Balmer lines signatures of the third cycle of our monitoring of omicron Ceti in $H\delta$ (full Stokes). See Figure \ref{fig:spec1}.}\label{fig:spec3}
\end{center}  
\end{figure*}

\begin{figure*}[h!]
\begin{center}
\includegraphics[scale=0.4]{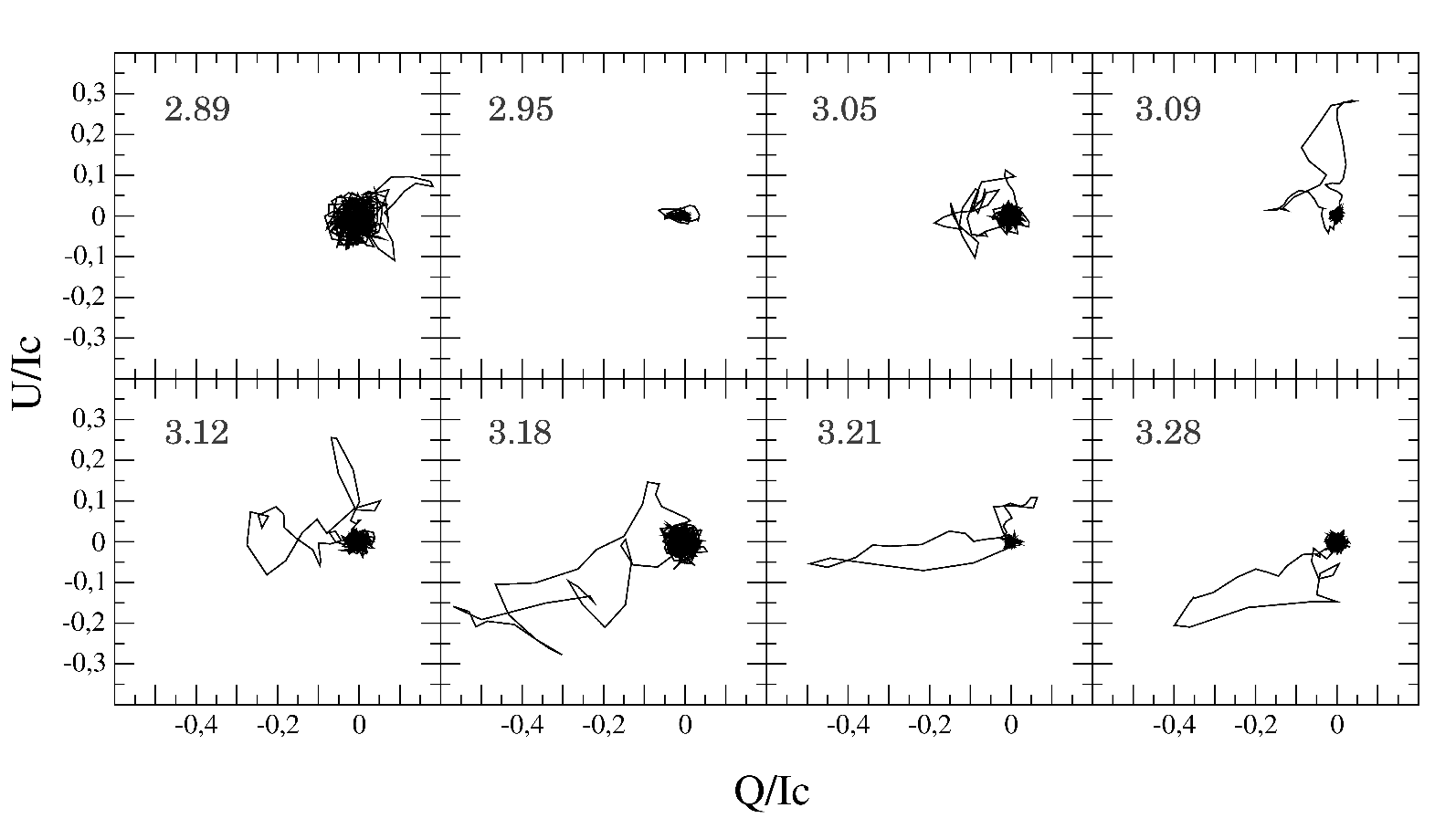}
\caption{Hydrogen Balmer lines signatures of the third cycle of our monitoring of omicron Ceti in $H\delta$ (u(q)). See Figure \ref{fig:spec2}.}\label{fig:spec4}
\end{center}  
\end{figure*}

From Figs \ref{fig:spec3} and \ref{fig:spec4}, which displays the data around the Balmer line $H\delta$, the intensity emission appears at $\varphi$=2.95 similar to what is observed for $H\gamma$. It starts to develop rapidly after the maximum light and continuously increases during the monitoring to reach a maximum amplitude  from $\varphi$=3.21 to  $\varphi$=3.28. The maximum intensity of $H\delta$  is much higher than for $H\gamma$. However, the temporal evolution of $H\delta$ is different in the sense that the intensity stays constant during the last two phases in contrast to what happens in $H\gamma$. This is probably because $H\delta$ photons are the most energetic. Since the shock is weaker at that time, they should be the ones whose production should start to be reduced first, and $H\delta$ is also less mutilated by molecular absorption than the $H\gamma$ line. One strong absorption at zero velocity severely affects the profile,  while another one is barely noticeable on the blue wing around maximum light.  The total width of the emission is constant through time at 75 \kms , which is smaller than the one in $H\gamma$. For this line, the width and position are barely modified through time.

As in $H\gamma$, from the maximum light up to the end of our observations, signatures are clearly detected in Stokes u, q, but also in v.
Stokes q varies from being slightly positive before the maximum in luminosity,  to nearly cancellation around $\varphi$=2.95, and then to negative values from $\varphi$=3.05. Then it starts to develop and is at a maximum from $\varphi$=3.18 until the end of our survey. The q signatures become narrow progressively and even  present a small peak appearing at some phases on the bluest side. Stokes u is undergoing dramatic changes. After being roughly zero until $\varphi$=3.05, it develops positively, getting as high as 0.3 at $\varphi$=3.09. Then this positive peak recedes while a negative structure develops, and finally  at $\varphi$=3.28 u signature is fully negative. Over a large portion of our monitoring, Stokes u presents  an "inverse P Cygni" structure, positive in the bluest part. This part of the spectrum corresponds to the one in intensity where the emission is the brightest. At last, the Stokes v parameter presents a very strong feature at phases $\varphi$=3.12 and 3.18, which almost disappears suddenly at following phases.

The evolution of the linear polarization angle in the u(q) plane is slightly similar to H$\gamma$. For phases such as $\varphi$=3.05 and $\varphi$=3.12,  the angle cannot be determined easily. However, we can definitely state that $\theta$ evolves from $45^{\circ}$ at $\varphi$=3.09 to eventually settle down at approximately $100^{\circ}$ at the last three observations (from $\varphi$=3.18 to $\varphi$=3.28). This is interesting when compared to H$\gamma$ where $\theta$ does not settle down. The linear polarization rate is at first as low as 0.1 at  $\varphi$=2.89 and clearer at $\varphi$=3.05. It increases up to 0.5 at $\varphi$=3.18 and 3.21 and then decreases to 0.4. This reduction in polarization seems to occur during the shock's weakening.

During those phases, Stokes v shows up before $\theta$ settles down to some value. We know that the decrease in the intensity emission is due to the deceleration of the shock wave. In \cite{gilletal1983}, it is shown that the emission line in $H\alpha$ increases until $\varphi$=0.40 (the reference maximum is of course different than ours) and then goes down. The change in the polarization behaviour described above thus certainly happens at the moment of this deceleration.

\subsubsection{Comparisons to previous cycles}

\begin{figure*}[h!]
\begin{center}
\includegraphics[scale=0.6]{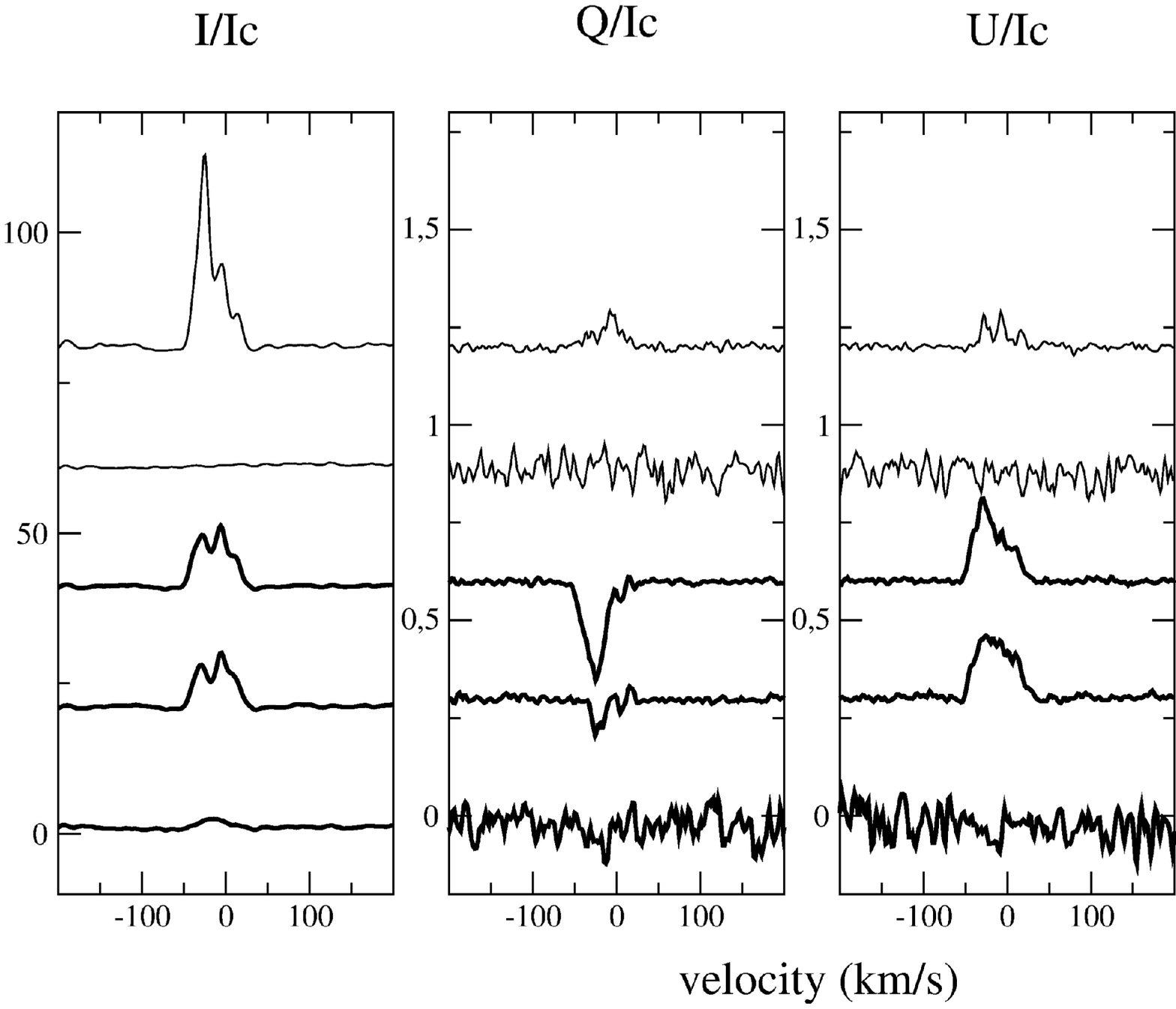}
\caption{Hydrogen Balmer lines signatures of cycles 1 (thick lines) and 2 (thin lines) of our monitoring of omicron Ceti in $H\gamma$ (full Stokes). See Figure \ref{fig:spec1}.}\label{fig:spec5}
\end{center}  
\end{figure*}

\begin{figure*}[h!]
\begin{center}
\includegraphics[scale=0.4]{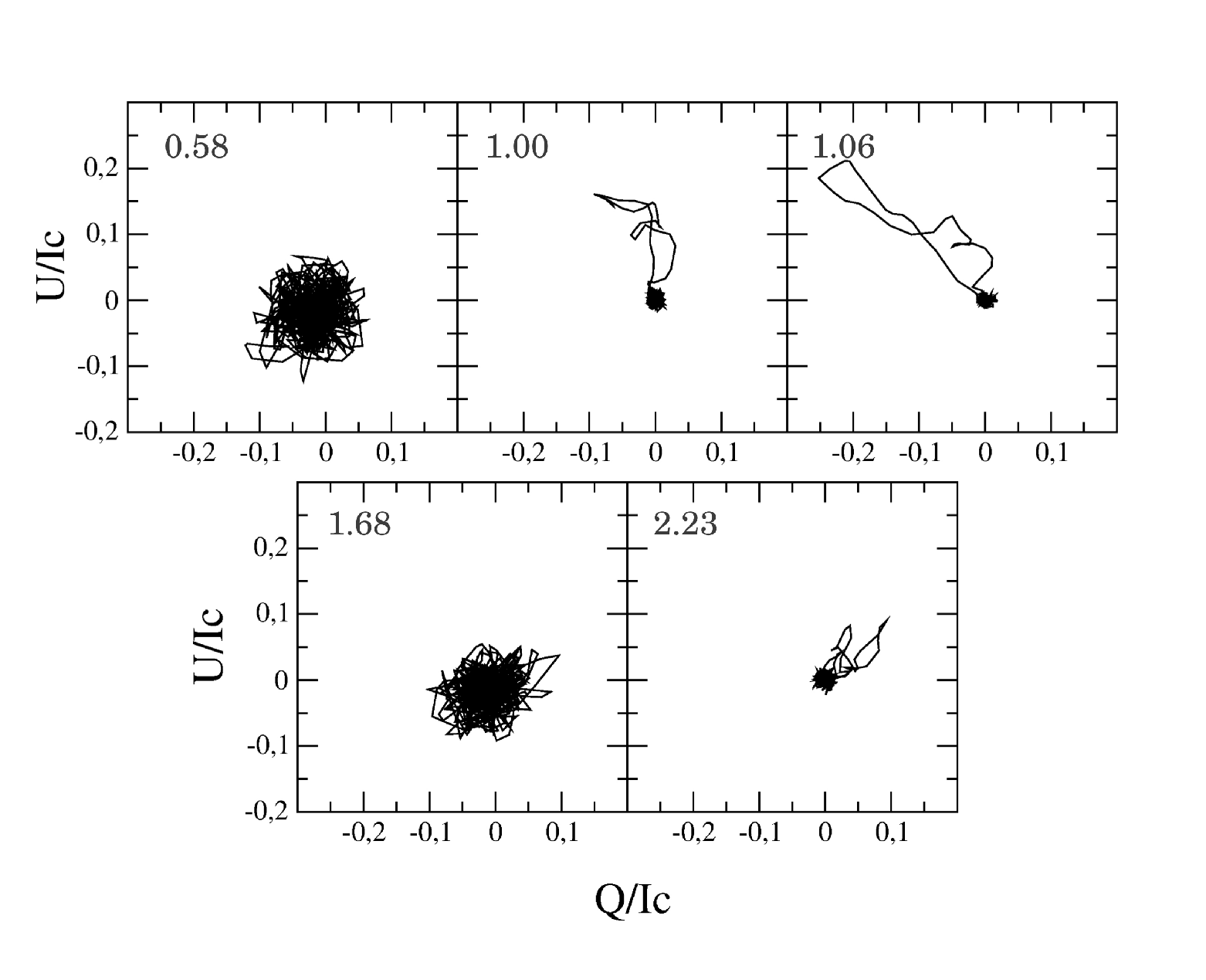}
\caption{Hydrogen Balmer lines signatures of cycles 1 and 2 of our monitoring of omicron Ceti in $H\gamma$ (u(q)). See Figure \ref{fig:spec2}.}\label{fig:spec6}
\end{center}  
\end{figure*}

\begin{figure*}[h!]
\begin{center}
\includegraphics[scale=0.6]{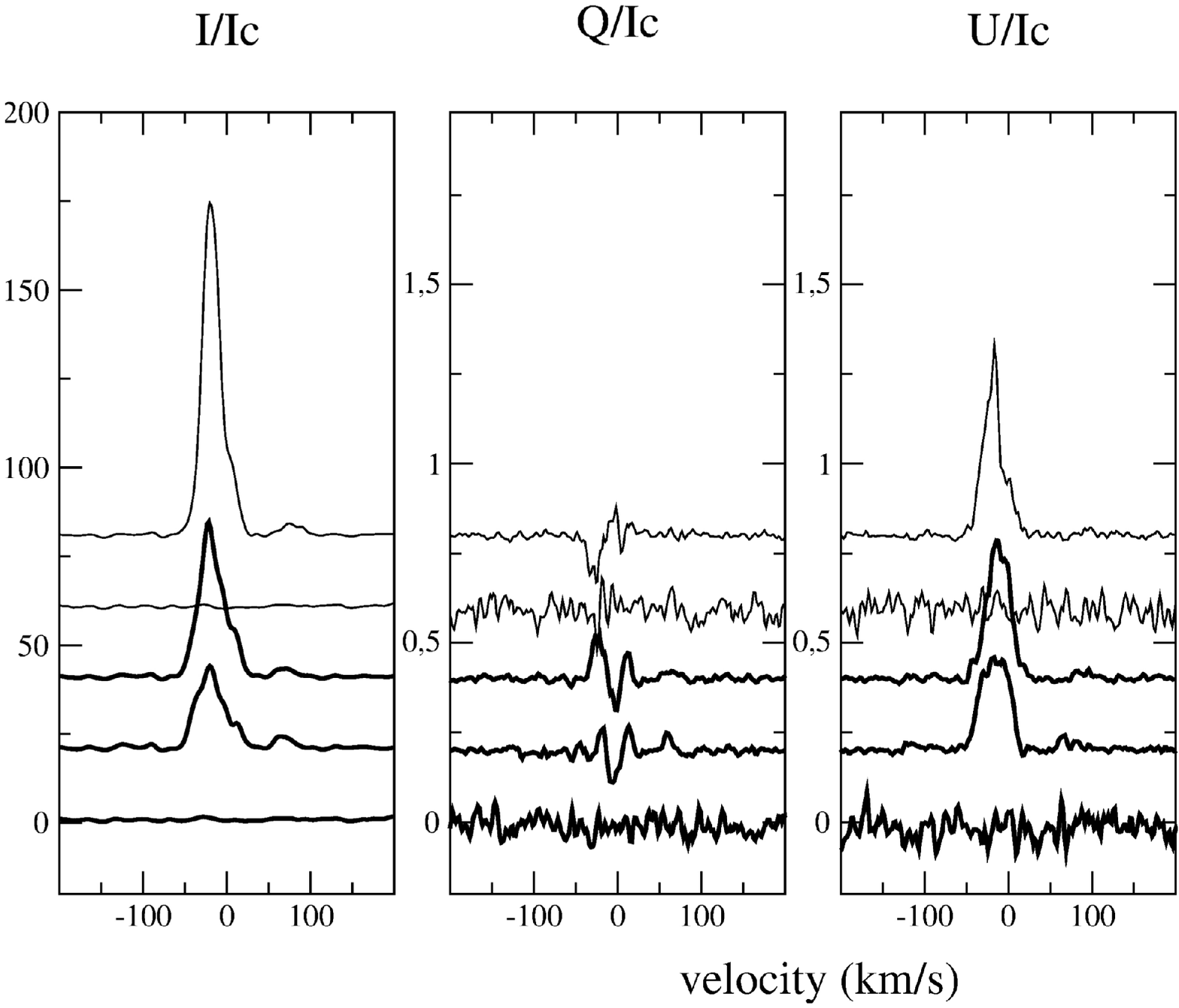}
\caption{Hydrogen Balmer lines signatures of cycles 1 (thick lines) and 2 (thin lines) of our monitoring of omicron Ceti in $H\delta$ (full Stokes). See Figure \ref{fig:spec1}.}\label{fig:spec7}
\end{center}  
\end{figure*}

\begin{figure*}[h!]
\begin{center}
\includegraphics[scale=0.4]{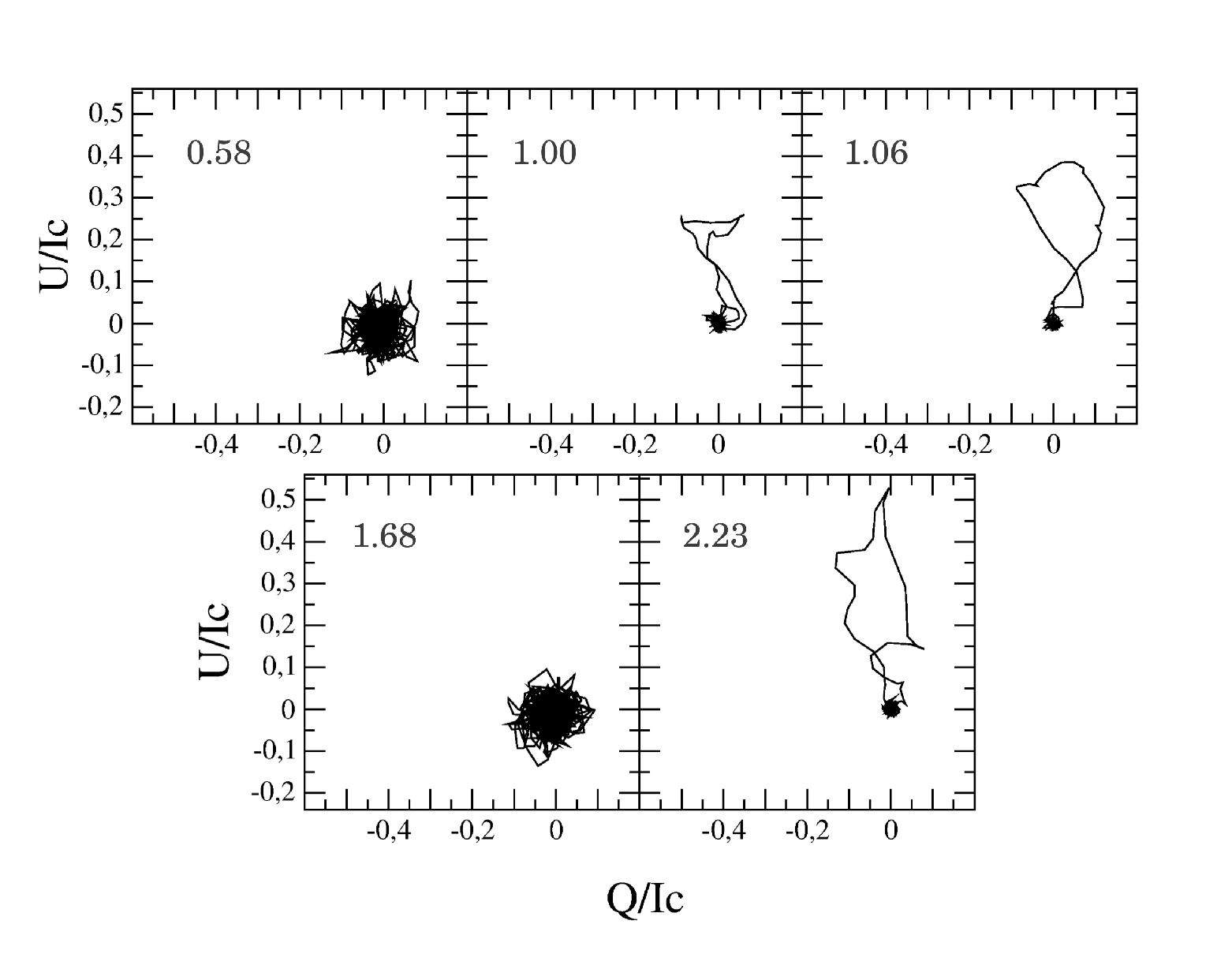}
\caption{Hydrogen Balmer lines signatures of cycles 1 and 2 of our monitoring of omicron Ceti in $H\delta$ (u(q)). See Figure \ref{fig:spec2}.}\label{fig:spec8}
\end{center}  
\end{figure*}

In Figures \ref{fig:spec5} to \ref{fig:spec8}, we present some phases from the previous, but less well-sampled cycles. When we compare the same phases of different cycles, the intensities of H$\gamma$ and H$\delta$  are roughly the same, in amplitude, width, and shape.  Therefore, we can safely state that, for the considered phases of each cycle, the shock wave will produce intensity emissions of similar strength.

In contrast, the u, q, and v Stokes parameters can strongly vary from one cycle to another. Both symmetries and amplitudes of q, u, and v signatures are found to be altered in many cases, inducing a change in the linear polarization angle.  Interestingly,  the linear polarization rate seems to be roughly conserved through cycles. This allows us to stress the non-repeatability of the polarizing processes in action. Indeed, we know already that each cycle of the luminosity variation of omicron Ceti is not an exact copy of the previous ones (see AAVSO website). Considering that the shock wave is the consequence of the pulsation of the star, as the luminosity variation is, then it is no wonder that the polarizing mechanism linked to the shock wave does not behave exactly the same as before. However, as we noticed for the third cycle, the signatures' widths do not change from one cycle to another for a given Balmer line. 

\section{Interpretation}
The appearance of an emission line indicates that the specific intensity of the back-lighting radiation (creating the continuum) is less than the source function inside the shock. This is linked to the intrinsic photon emission behind the shock's front. Emission lines can thus help in determining the properties of this area. Polarization also happens to be much stronger in the Balmer lines than in the pseudo-continuum around them. Its temporal evolution follows the evolution of intensity emission. It is therefore believed that the polarizing mechanism is also intrinsic to the shock wave.

In Figures \ref{fig:spec1} to \ref{fig:spec4}, the evolution of intensity emission lines H$\delta$ and H$\gamma$  is quite similar, leading us to think that the shock emerges from the photosphere just before the maximum light and is getting stronger toward the end of the third cycle of our survey. For both Balmer lines, the emission's width and central velocity (expressed in stellar rest frame) do not vary with time as previously said. This would imply that the shock is not accelerating much. 
Moreover, if the width of the emission remains the same, it probably means that the mechanism behind line broadening (micro-turbulence?) does not evolve with time.
\cite{will1976} already addressed the characterization of shock wave properties through emission lines width in Miras. In the case of an optically thin medium and when broadening line mechanisms are not dominant, the width is an indicator of the flow velocity where the radiation is emitted. When the gas is optically thick, we cannot give anything else but an upper limit for this velocity. Moreover, the model of shock wave we chose also influences this relationship. In \cite{fadgil2004}, shock waves models are used to create synthetic Balmer lines with different values of temperature, density and upstream velocity. The models show a line doubling that is particularly obvious for $H\alpha$ and just slightly obvious for $H\beta$ while $H\gamma$ stays single-peaked. On the other hand, the widths of the emissions vary clearly with density and upstream velocity.

A few shock-related mechanisms of polarization are probably operating. First of all, we consider a photospheric origin. The shock starts deep in the atmosphere, probably even below the photosphere. The photospheric origin has already been thought of for polarization (e.g. \citealt{harrington1969}) but never with a shock wave crossing it. Besides this, it is known that stars with low gravity, such as Mira stars, have large convective cells. In \cite{Chiavassa09}, 3D models of red supergiants (RSG) photosphere were compared to interferometric observations in the near infrared. They conclude that RSG have large-scale granulations undergoing a temporal variation. This kind of work has also been done for AGB stars (\citealt{freytaghofner08}). Thus, this convective structure could transfer its asymmetry to the propagating shock wave. Since the convective cells are very large, the asymmetry is not cancelled by integration on the stellar disk, and the shock wave may be aspherical. However, other sources of asymmetry (e.g. high-layer asymmetries) might be of interest in this respect (\ca{rag08}, \ca{pluz09}, \ca{chia10}, \ca{witt11}). \cite{Fadeyev01} have shown that the compression rate $\eta$ behind a radiative shock can be written as

\begin{equation}
\eta=4+3\frac{E_{in}-E_{in1}}{E_{t}}+3\frac{F_R-F_{R1}}{\rho{U}E_t}
\end{equation}

\noindent where $E_{t}$ is the specific energy in the translational degrees of freedom, $E_{in}$ the specific energy of excitation and ionization of atoms, and $F_R$ the radiation flux. The subscript 1 refers to the values in the state ahead of the shock. Thus in the presence of radiation, the gas compression ratio can reach values of tens or even hundreds, significantly greater than the compression ratio without radiation ($\eta=4$). With such high compression ratios, the level of turbulence of the gas must be strongly amplified (see \citealp{Gillet98_2}). When the shock wave passes through the convection zone located around the photosphere, the convective cells are thus sharply compressed and heated. from the point of view of the shock, the convective structure is almost static. It is likely that the convective field is disrupted and even destroyed temporarily during the passage of the wave. Consequently, if the heterogeneous distribution of convective cells causes polarization, then it should also be observed in the continuum and be disturbed during the photospheric passage of shock. It seems that the quality and temporal resolution of current observations are not yet sufficient to confirm this point directly. New observations are needed.

More local phenomena can be called upon. When the Mach number of shock wave increases, its structure quickly becomes complex. In particular, the charge separation within the shock front region may induce an electric field. The nature of the resultant interaction of the electric field with the gas flow has been investigated for a steady plane shock in fully and partially ionized gases (see for instance \citealp{Jaffrin65}, \citealp{Lu74}, \citealp{Vidal95}). The electric field is parallel to the direction of shock propagation. However, the presence of background density or temperature gradients, for instance, causes an electrical current, inducing a curl of magnetic field. Thus the shockfront becomes a source of magnetic field, which pervades the unperturbed gas and the wake of the shock. These studies were limited to the region of viscous shock and did not consider the plasma state appearing later in the wake when the coupling with the radiation field can be fundamental. As discussed above, the solenoidal state of the postshock gas strongly increases in the presence of a radiative field (shock turbulence amplification, see \citealp{Gillet98_2}). Consequently, because the shock propagates in a non-uniform background plasma, a spontaneous generation of a magnetic field is possible within the postshock plasma. From the induction equation, the equation describing the development of magnetic field is

\begin{equation}
\frac{\partial{B}}{\partial{t}}=\nabla\times(v\times{B})+\eta\nabla^2B+S.
\end{equation}

\noindent The three terms on the right-hand side describe the convection, diffusion, and generation of a magnetic field. The self-generation magnetic field could be due to several local mechanisms such as ion-electron separation at the front or density and temperature gradients in the plasma. Thus the dominant physical process must be determined to write the source term S. The magnetic field is perpendicular to the shock propagation direction. Consequently, a linear polarization could be expected, although because of the complexity of the environment, a minor circular polarization could also be observed.

The observation of the polarization of emission lines has been a long-standing issue in solar flare physics (\citealp{haug71}), in tokamak (\citealp{fujimoto96}), gasdischarge (\citealp{walden99}), and laser produced plasmas (\citealp{yoneda97}). The emission polarization can come from different causes induced by the non-uniform background plasma. In particular, it has been known that anisotropic collisions of electrons with atoms produce polarized radiations in plasma in which the collisional excitation and ionization processes dominate (impact polarization). Thus, the generation of polarized emission can occur during the recombination processes. In fact, there are few theoretical and experimental works about this matter, and devoted to radiation fronts, laboratory ionizing shock tubes, or radiative shock waves. In conclusion, we expect that the observed polarization in emission lines of Mira could have been caused by the spatial non-uniformity in the electron distribution function that was induced by the self-generated magnetic field created in the shockfront. It is not possible within the framework of this study to definitively explore the physical origin of this polarization. Only a quantitative approach may allow us to answer.

Tables \ref{tab2} and \ref{tab3} sum up the evolution of spectropolarimetric values for H$\gamma$ and H$\delta$ for the 3rd cycle. Considering only I profiles, it is difficult to state at which moment of the cycle the shock reaches its maximum intensity because the emissions do not undergo the same molecular absorption. Then, polarization could help determine this moment. If we consider the local mechanism, the turbulence would be at its maximum during phases of most intense emissions, inducing the strongest polarization. It is interesting to see that linear polarization appears during a longer time than the circular one. Stokes V is rarer and mostly present when the linear polarization is quite strong, i.e. at about phase 0.20. Indeed, this is consistent because when the shock becomes strong, the turbulence amplification is at a maximum, and consequently the hydrodynamic motions are more and more tridimensional. It is interesting to note that phases around 0.20 are quite common in atmospheric models for merging shock fronts (\ca{ireland11}).

\begin{table}[h]
\caption{Time evolution of spectropolarimetric values in H$\gamma$.}
\label{tab2}
\begin{minipage}{\linewidth}
\begin{tabular}{|c|c|c|c|c|c|c|c|c|c|}
\hline
$\varphi\footnote{The phase along our survey of the third cycle.}$ & 2.89 & 2.95 & 3.05 & 3.09 & 3.12 & 3.18 & 3.21 & 3.28 \\
\hline
I/Ic\footnote{From lines 2 to 6 are the maximum values of the indicated parameter.}& 0 & 6 & 12 & 14 & 14 & 15 & 20 & 26\\
\hline

 Q/Ic & 0.1 & 0 & 0.1 & 0.05 & -0.1 & -0.2 & -0.15 & -0.15 \\
\hline
 U/Ic & 0 & 0& 0.2& 0.22&0.2 & 0.15&0.2 &0 \\
\hline
 V/Ic &0 &0 &0 &0 &0 &0 &0 &-0.1 \\
\hline
 p$_{lin}$/Ic & 0.1& 0&0.2 &0.22 &0.2 &0.25 &0.22 &0.15 \\
\hline
 $\theta$ &20$^{\circ}$&0$^{\circ}$ &40$^{\circ}$ &45$^{\circ}$ &45$^{\circ}$ &80$^{\circ}$ &80$^{\circ}$ & 90$^{\circ}$\\

\hline
\end{tabular}
\end{minipage}
\end{table}

\begin{table}[h]
\caption{Time evolution of spectropolarimetric values in H$\delta$ (see Table \ref{tab2}).  }
\label{tab3}
\begin{tabular}{|c|c|c|c|c|c|c|c|c|c|}
\hline
$\varphi$ & 2.89 & 2.95 & 3.05 & 3.09 & 3.12 & 3.18 & 3.21 & 3.28 \\
\hline
I/Ic & 5 & 8 & 38 & 55 & 60 & 65 & 80 & 80\\
\hline

 Q/Ic & 0.15 & 0 & -0.15 & -0.15 & -0.25 & -0.5 & -0.5 & -0.4 \\
\hline
 U/Ic & 0 & 0& 0.1 & 0.25 & 0.2 & -0.2 &0.1 &-0.2 \\
\hline
 V/Ic &0 &0 &-0.1 & -0.05 & -0.26  &-0.5 & 0 &-0.1 \\
\hline
 p$_{lin}$/Ic & 0.15& 0&0.15 &0.28 &0.25 &0.5 &0.5 &0.4 \\
\hline
 $\theta$ &20$^{\circ}$ &0$^{\circ}$ &90$^{\circ}$ &45$^{\circ}$ &90$^{\circ}$ & 100$^{\circ}$ &90$^{\circ}$ & 100$^{\circ}$\\

\hline
\end{tabular}
\end{table}

To know which of those two phenomena are likely to occur, it is necessary to make similar observations of other types of radially pulsating stars that have no convection, but are known to host an atmospheric shock wave. The detection of polarization in the Balmer emission lines of such stars would then give credit to the local hypothesis. On the other hand, if the polarization arises from the convective asymmetry, it would also act on the continuum of the photosphere's radiation. At that point, the global hypothesis would prevail. A temporal survey of the polarization of this continuum would then be necessary to see whether it varies like the Balmer lines polarization presented in Tables \ref{tab2} and \ref{tab3}.

\section{Conclusion}

We performed the first spectropolarimetric monitoring of the Mira star omicron Ceti, focussing on Balmer emission lines. Thanks to these observations, we were able to detect variable linear and circular polarization signatures associated to those hydrogen lines. The temporal variation of those signatures is correlated with the evolution of the intensity of the emission lines known to be formed in the radiative wake of a shock wave propagating throughout omicron Ceti's atmosphere.

Several potential polarizing mechanisms are considered. AGB stars host very large photospheric convective cells inducing overhall asymmetry. 
Since the shock is formed near the photosphere, it is possible for it to interact with the convective structure and thus to gain the photospheric asymmetry. Interferometric observations at minimum light, i.e. when the shock is high in the atmosphere, would help to compare the photosphere's and the shock's geometrical features. On the other hand, our local hypothesis states that a charge separation induced by the shock's passage along with the turbulence believed to exist behind the front of the shock could produce a local magnetic field through which the radiation from the wake becomes polarized.

Unlike omicron Ceti, other Mira stars can be rich in carbon or have no companion. They would thus be interesting to observe to determine how general this evolution of the line polarization is. This study might also be extended to other kinds of pulsating variables stars, such as RV Tauri or RR Lyrae, known to host radiative shock waves propagating throughout the atmosphere. If pulsating stars with no convection (such as $\beta$ CMa stars) were still showing such line polarization, it would favour the local hypothesis.

\section{Acknowledgments}
We acknowledge with thanks the variable star observations of the AAVSO International Database contributed by observers worldwide and used in this research. We also acknowledge financial support from the "Programme National de Physique Stellaire" (PNPS) of CNRS/INSU, France. We are grateful to V. Bommier, A. Chiavassa, J.-F. Donati, J. Morin, and F. Paletou for useful comments.
\bibliography{bibfabas}

\end{document}